\documentclass[11pt,twoside]{article}
\usepackage{graphicx}
\usepackage{amsmath}
\usepackage{amssymb}
\usepackage{cite}

\begin{document}
\newcommand{\pst}{\hspace*{1.5em}}

\newcommand{\rigmark}{\em Journal of Russian Laser Research}

\newcommand{\be}{\begin{equation}}
\newcommand{\ee}{\end{equation}}
\newcommand{\bm}{\boldmath}
\newcommand{\ds}{\displaystyle}
\newcommand{\bea}{\begin{eqnarray}}
\newcommand{\eea}{\end{eqnarray}}
\newcommand{\ba}{\begin{array}}
\newcommand{\ea}{\end{array}}
\newcommand{\arcsinh}{\mathop{\rm arcsinh}\nolimits}
\newcommand{\arctanh}{\mathop{\rm arctanh}\nolimits}
\newcommand{\bc}{\begin{center}}
\newcommand{\ec}{\end{center}}
\newcommand{\ket}[1]{{\left| #1 \right>}}
\newcommand{\bra}[1]{{\left< #1 \right|}}
\newcommand{\aver}[1]{{\left< #1 \right>}}

\thispagestyle{plain}

\label{sh}


\begin{center} {\Large \bf
\begin{tabular}{c}
CHAOTIC WALKING AND FRACTAL SCATTERING OF ATOMS \\
IN A TILTED OPTICAL LATTICE
\end{tabular}
 } \end{center}

\bigskip

\bigskip

\begin{center} {\bf S.V.~Prants, V.O. Vitkovsky}
\end{center}

\medskip

\begin{center}
{\it
Laboratory of Nonlinear Dynamical Systems,\\
Pacific Oceanological Institute of the Russian Academy of Sciences,\\
690041 Vladivostok, Russia, URL: dynalab.poi.dvo.ru}

\smallskip

$^*$Corresponding author e-mail:~~~prants@poi.dvo.ru
\\
\end{center}
\begin{abstract}\noindent
Chaotic walking of cold atoms in a tilted optical 
lattice, created by two counter propagating running waves with an additional 
external field, is demonstrated theoretically and numerically in the semiclassical 
and Hamiltonian approximations. 
The effect consists in random-like changing the direction of atomic motion 
in a rigid lattice under the influence of a constant force due to a specific behavior of the atomic 
dipole-moment component that changes abruptly in a random-like manner 
while atoms cross standing-wave nodes. Chaotic walking generates 
a fractal-like scattering of atoms that manifests itself in a self-similar 
structure of the scattering function in the atom-field detuning, position and momentum spaces. 
The probability distribution function of the scattering time is shown to 
decay in a non-exponential way with a power-law tail.    
 
\end{abstract}

\medskip


\section{Introduction}
\pst

The mechanical action of light upon neutral atoms placed in a laser 
standing wave  is at the heart of laser cooling, trapping, and Bose-Einstein
condensation \cite{RMP}. Numerous applications of the mechanical action of light
include isotope separation, atomic interferometry, atomic lithography and epitaxy,
atomic-beam deflection and splitting, manipulating
translational and internal atomic states, measurement of atomic positions, 
etc. Atoms and ions in an optical lattice, formed by a laser standing wave, are perspective objects for
implementation of quantum information processing and
quantum computing. Advances in
cooling and trapping of atoms, tailoring optical potentials
of a desired form and dimension, controlling the level of dissipation
and noise are now enabling the direct experiments
with single atoms to study fundamental
principles of quantum physics, quantum chaos, decoherence,
and quantum-classical correspondence.

Nonlinear dynamics of cold atoms in optical lattices is a fastly growing 
branch of atomic physics. There are a number of theoretical works and impressive 
experiments on quantum chaos, dynamical localization, chaos-assisted tunneling, 
L\'evy flights, etc. (for reviews see \cite{Raizen,Pbook}). To suppress spontaneous 
emission and provide a coherent quantum dynamics one usually works far
from the optical resonance. Adiabatic elimination of
the excited state amplitude leads to an effective
Hamiltonian for the center-of-mass motion, whose 3/2 degree-of-freedom classical analogue
has a mixed phase space with regular islands embedded in a chaotic sea.
New possibilities are opened if one works near the optical resonance 
and take the internal atomic dynamics into account. A single atom in a standing-wave
laser field may be semiclassically treated as a nonlinear dynamical
system with coupled internal (electronic) and external (mechanical) degrees
of freedom \cite{JETPL01,PRA01,JETPL02}. In the semiclassical
and Hamiltonian limits (when one treats atoms as point-like particles and neglects
spontaneous emission and other losses of energy), a number of nonlinear dynamical
effects have been analytically and numerically demonstrated
with this system: chaotic Rabi oscillations
\cite{JETPL01,PRA01,JETPL02}, Hamiltonian chaotic atomic
transport and dynamical fractals \cite{JETP03,PLA03,PU06,PRA07}, 
L\'evy flights and anomalous diffusion
\cite{JETPL02,PRE02,JRLR06}. These effects are caused by
local instability of the CM motion in a laser
field. A set of atomic trajectories under certain conditions becomes exponentially sensitive to small
variations in initial quantum internal and classical external states
or/and in the control parameters, mainly, the atom-laser detuning. Hamiltonian evolution
is a smooth process that is well described in a semiclassical approximation by the coupled Hamilton-Schr\"odinger
equations. A detailed theory of Hamiltonian and dissipative chaotic transport of
atoms in a laser standing wave has been developed in Refs.~\cite{PRA07} and 
\cite{PRA08,EPL08}, respectively.

Additional possibilities to manipulate the atomic transport are created 
by applying an external force to the standing-wave optical potential. 
It is obvious that for cold atoms in a vertical optical lattice it is necessary 
to account for the Earth's acceleration. It is possible as well to create 
horizontal accelerated optical lattices by adding a constant force 
whose magnitude along the optical axis can be 
easily varied. The problem of atomic motion in a tilted optical potential is closely related to the old 
problem of electron motion in a a periodic crystal with dc or ac forces applied. 
The analogue of well known Bloch oscillations with cold atoms has been 
experimentally found in Ref.\cite{Ben,Fischer}.

In the present paper we apply the ideas and methods, elaborated in the field of 
nonlinear dynamics of cold atoms, to study 
theoretically and numerically motion of point-like atoms in a tilted 
optical lattice. It will be shown that varying only one parameter, 
the detuning between the frequencies of a working atomic transition and 
the laser field, one can explore a variety of regimes of atom motion, 
including chaotic walking, dynamical fractals and chaotic scattering. 

\section{Chaotic and regular regimes of motion of atoms in a tilted potential}
\pst

In the one-dimensional case, the Hamiltonian of a two-level atom in a  
standing-wave laser field and an additional external field can be written 
in the frame rotating with the laser frequency $\omega_f$ as follows:
\begin{equation}
\begin{gathered}
\hat H=\frac{P^2}{2m_a}+\frac{\hbar}{2}(\omega_a-\omega_f)\hat\sigma_z-
\hbar \Omega_0 \left(\hat\sigma_-+\hat\sigma_+\right)\cos{k_f X} + FX,
\end{gathered}
\label{Ham}
\end{equation}
where $\hat\sigma_{\pm, z}$ are the Pauli operators for the internal atomic degrees
of freedom, $X$ and $P$ are the classical atomic position and momentum, 
$\omega_a$ and $\Omega_0$ are the atomic transition and maximal Rabi 
frequencies, respectively. $F$ stands for the static force induced by 
external field.

In the semiclassical approximation, where the transversal atomic momentum $p$ is 
supposed to be, in average, much larger than the photon one $\hbar k_f$, atom with quantized internal dynamics is  
treated as a point-like particle to be described by the Bloch--Hamilton  
equations of motion without relaxation terms 
\begin{equation}
\begin{gathered}
\dot x=\omega_r p,\quad \dot p=- u\sin x - \kappa, \quad \dot u=\Delta v,
\\
\dot v=-\Delta u+2 z\cos x, \quad
\dot z=-2 v\cos x, 
\end{gathered}
\label{Hamsys}
\end{equation}
where $u$ and $v$ are synchronized (with the laser field) and quadrature 
components of the atomic electric dipole moment, respectively, and $z$ is 
the atomic population inversion. Equations (\ref{Hamsys}) are written in
the dimensionless form with $x\equiv k_f X$ and $p\equiv P/\hbar k_f$ to be classical
atomic center-of-mass position and momentum, respectively.
Dot denotes differentiation with respect to the dimensionless time $\tau\equiv \Omega t$.
The set (\ref{Hamsys}) has the three control parameters
\begin{equation}
\omega_r\equiv\hbar k_f^2/m_a\Omega, \quad
\Delta\equiv(\omega_f-\omega_a)/\Omega, \quad \kappa \equiv F/\hbar k_f\Omega, 
\label{control}
\end{equation}
which are the normalized recoil frequency, $\omega_r$,
atom-field detuning, $\Delta$, and applied force $\kappa$.
The system has two integrals of motion, namely the total energy
\begin{equation}
H\equiv\frac{\omega_r}{2}p^2 + \kappa x - u\cos x-\frac{\Delta}{2}z,
\label{H}
\end{equation}
and the length of the Bloch vector $u^2+v^2+z^2=1$.

The external force is directed in the negative direction of the optical axis $x$. 
So, if the initial atomic momentum, $p_0$, is chosen to be in the negative direction, 
the force will simply accelerate the corresponding atoms. If $p_0>0$, 
then one may expect much more complicated motion. 

Equations (\ref{Hamsys}) constitute a nonlinear Hamiltonian
autonomous system with two and half degrees of freedom. Owing 
to two integrals of motion, phase points move on a three-dimensional
hypersurface with a given energy value $H$. In general, motion in 
a three-dimensional phase space in characterized by a positive
Lyapunov exponent, a negative exponent equal in magnitude 
to the positive one, and zero exponent.
The maximal Lyapunov exponent characterizes the mean rate of the
exponential divergence of initially close trajectories and serves as 
a quantitative measure of dynamical chaos in the system.
Because of a transient character of chaos, we have computed  
the finite-time Lyapunov exponent $\lambda$ 
by the method developed in Refs.~\cite{JMP96,KP97}. It has been found that 
at the fixed value, $\omega_r=10^{-3}$, of the recoil frequency 
$\lambda$ is positive in the following ranges of  
values of the other control parameters: the detuning $-0.5 < \Delta <0.5$ and 
the force $-0.25 < \kappa <0.6$. Therefore, we expect chaotic motion of atoms 
with the parameter's values in those ranges. 

In numerical experiments throughout the paper we suppose that two-level atoms 
are initially prepared in the ground states, $u_0=v_0=0,z_0=-1$, with $x_0=0$ and 
fixed values of the two control parameters, the   
normalized recoil frequency, $\omega_r=10^{-3}$, and the external force
$\kappa =0.01$. The atom-field detuning, $\Delta$, can be changed in a wide range.
It will be shown in this paper 
that atoms may perform chaotic walking, the 
new type of motion in absolutely deterministic environment  
where atoms can change the direction of motion alternating between flying through 
the standing wave and being trapped in its potential wells. 
We would like to stress that the local instability produces chaotic center-of-mass 
motion in a rigid optical lattice without any modulation of its parameters.

\begin{figure}[!tpb]
\includegraphics[width=0.3\textwidth]{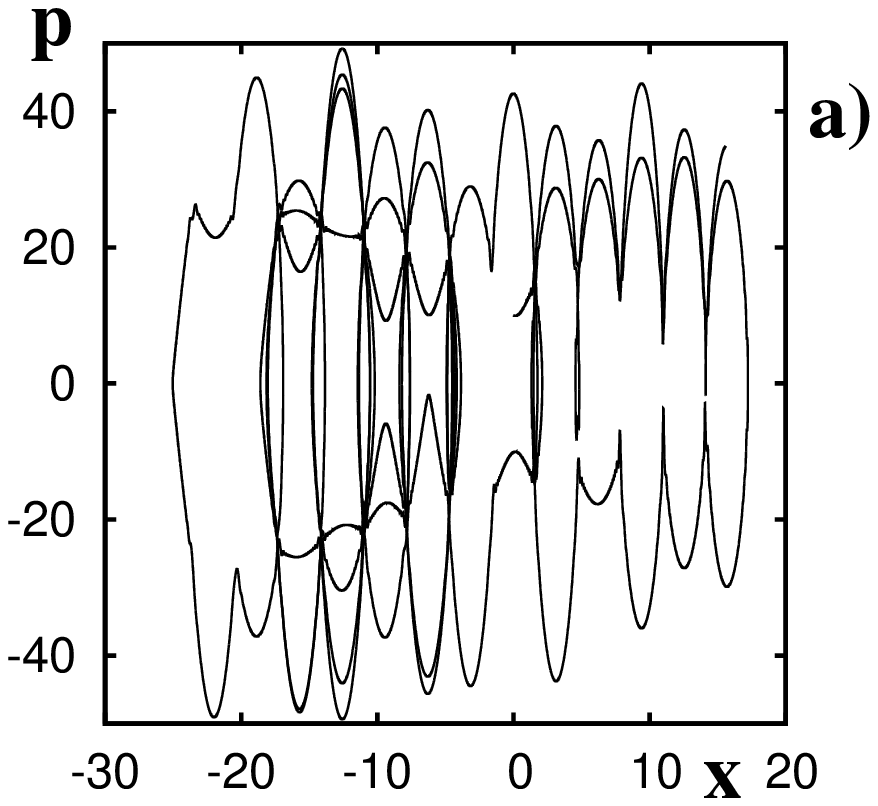}
\includegraphics[width=0.3\textwidth]{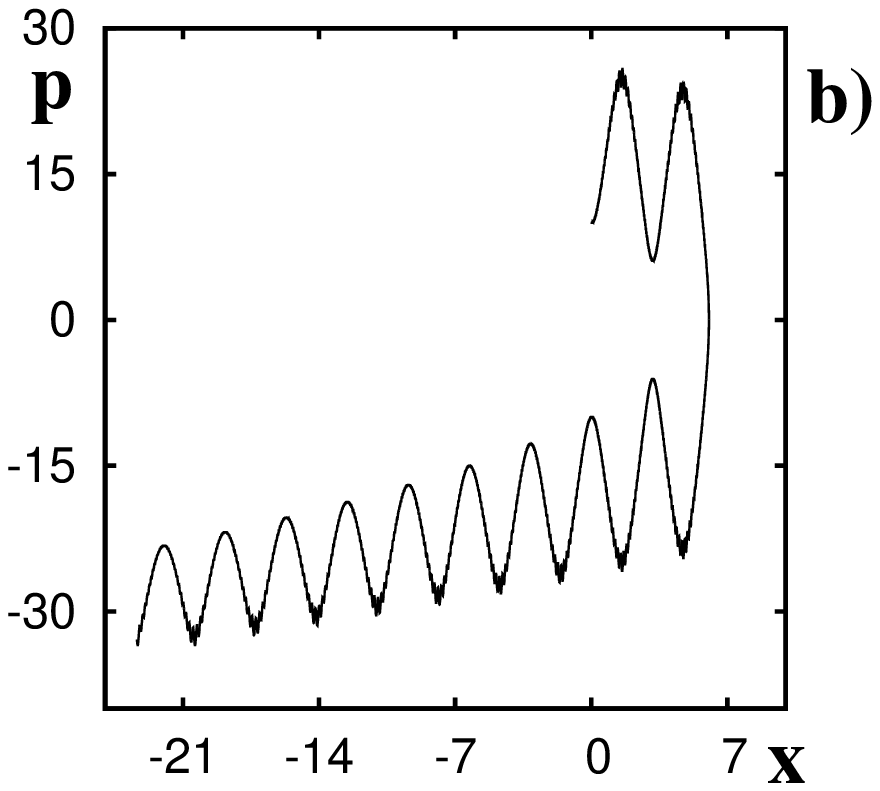}
\includegraphics[width=0.3\textwidth]{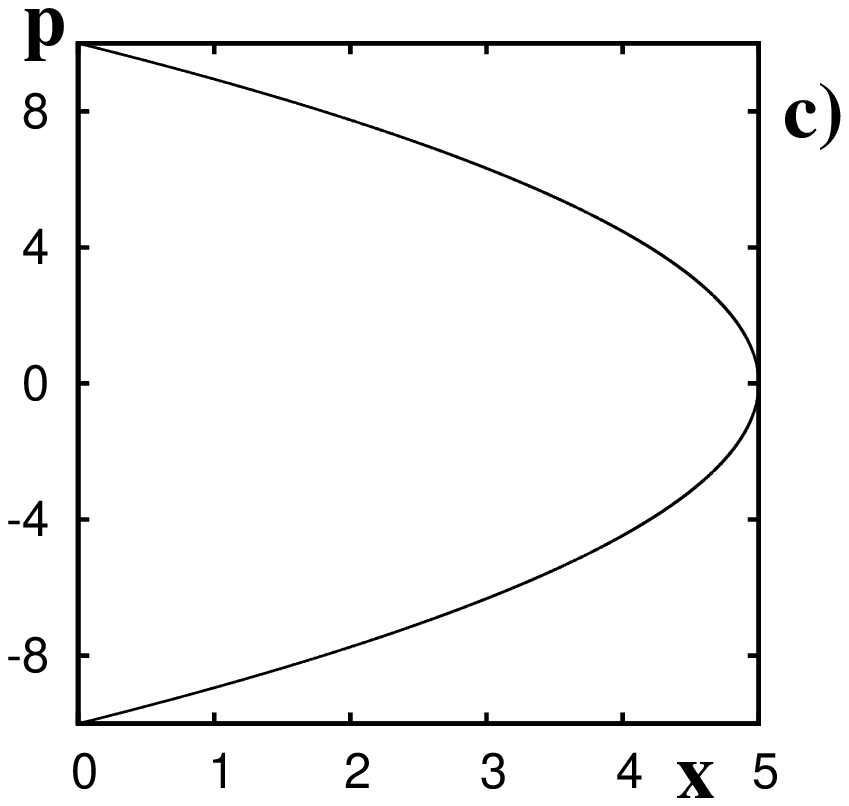}
\caption{Motion of a cold atom in a deterministic tilted optical 
lattice as it looks on the phase plane $x-p$. (a) Chaotic walking at 
$\Delta=0.15$, $\kappa=0.01$, $\omega_r=10^{-3}$. 
(b) Regular motion at $\Delta=1$ and with the same other conditions. 
(c) Regular motion at the resonance, $\Delta=0$, with the same other conditions.}  
\label{fig1}
\end{figure}

In Fig.~\ref{fig1} we illustrate different regimes of the center-of-mass 
atomic motion along the optical axis with the initial atomic momentum chosen to be 
$p_0=10$. It is simply the motion on the phase 
plane $x-p$. A typical picture with chaotic walking is shown 
in Fig.~\ref{fig1}a with the value of the detuning, $\Delta=0.15$, at which 
the maximal Lyapunov exponent, $\lambda$, is positive. The atom starts to move 
in the positive $x$-direction, changes soon the direction of motion a few times, 
acquiring irregularly positive and negative values of the momentum, and suddenly 
begins to move in the positive $x$-direction for a comparatively long time. 
Then it is decelerated, turns back  and flies in the negative 
direction. After that it changes the direction of motion many times 
demonstrating what we call ``chaotic walking''. 

For comparison, we show in Fig.~\ref{fig1}b the phase-plane 
motion with the larger detuning $\Delta=1$ (and at the same other conditions) 
at which the maximal Lyapunov exponent is 
not positive. The atom moves initially in the positive $x$-direction, accelerating and 
decelerating alternatively. Soon it changes the direction of motion and 
moves permanently in the negative direction.  
The motion is regular with a slight modulation of the momentum. 
The motion at exact resonance, $\Delta=0$, is even more simple 
(Fig.~\ref{fig1}c).   

\begin{figure}[!tpb]
\includegraphics[width=0.3\textwidth]{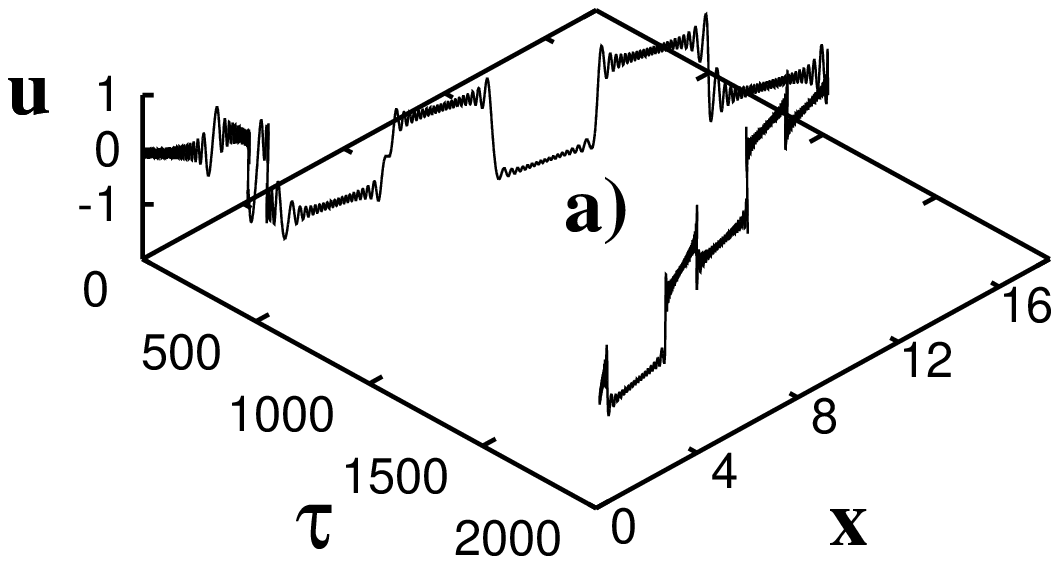}
\includegraphics[width=0.3\textwidth]{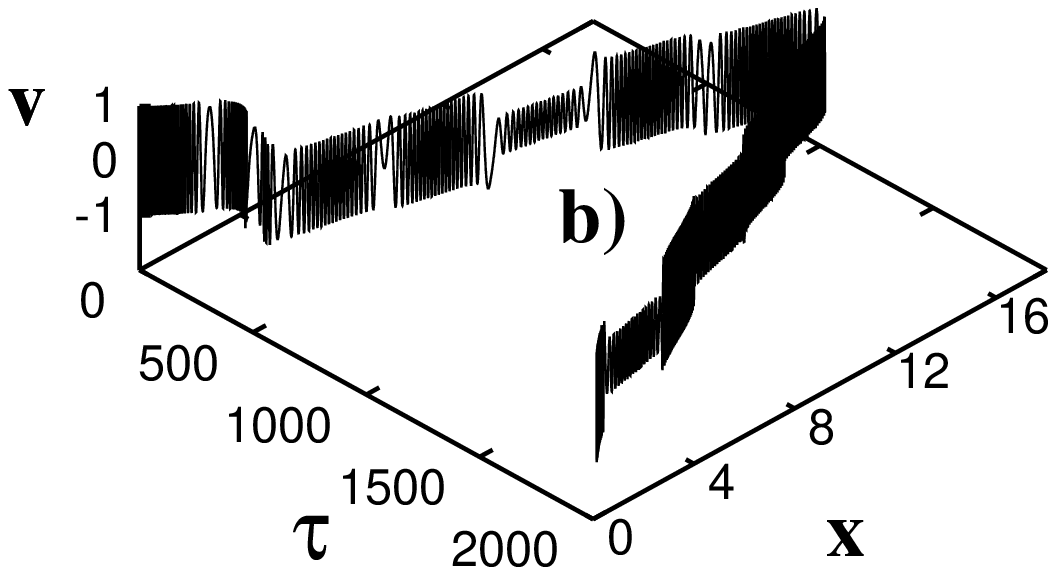}
\includegraphics[width=0.3\textwidth]{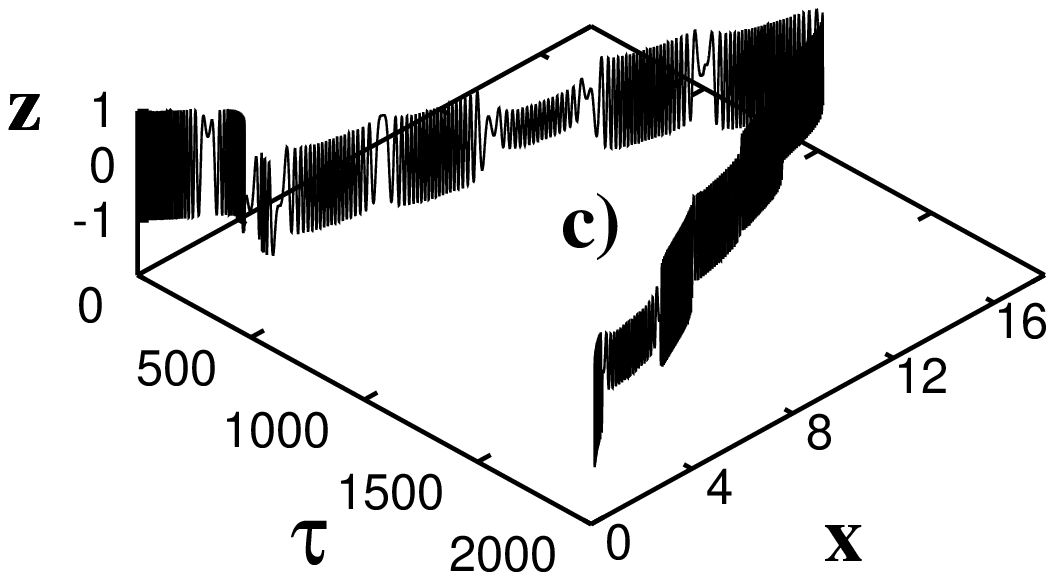}
\caption{Evolution of the atomic Bloch components in the regime of 
chaotic walking ($\Delta=0.15$).}
\label{fig2}
\end{figure}

\begin{figure}[!tpb]
\includegraphics[width=0.3\textwidth]{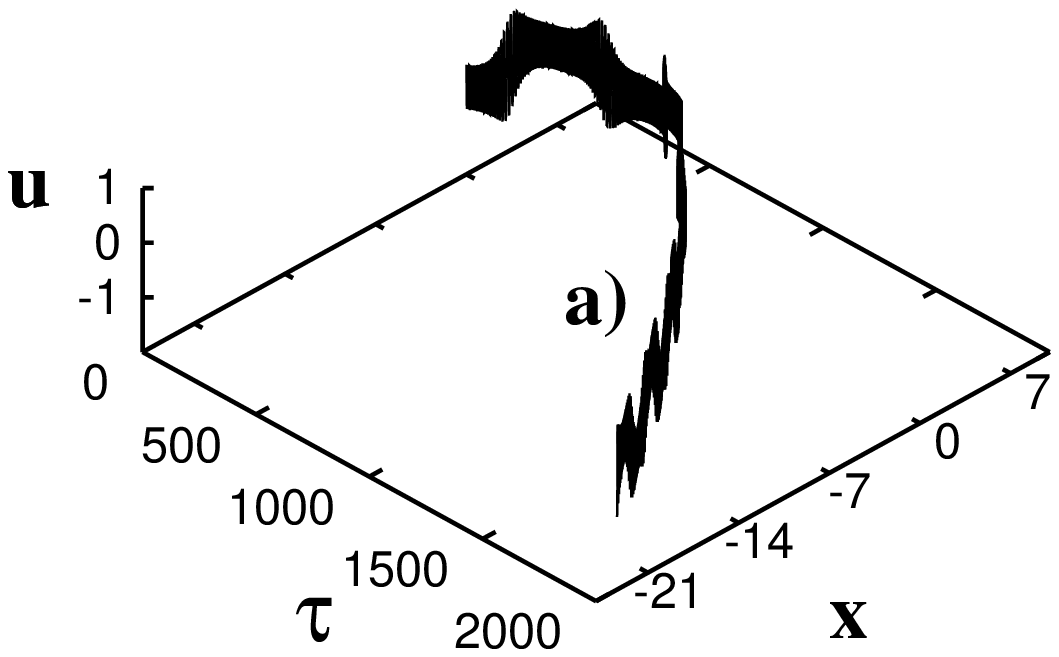}
\includegraphics[width=0.3\textwidth]{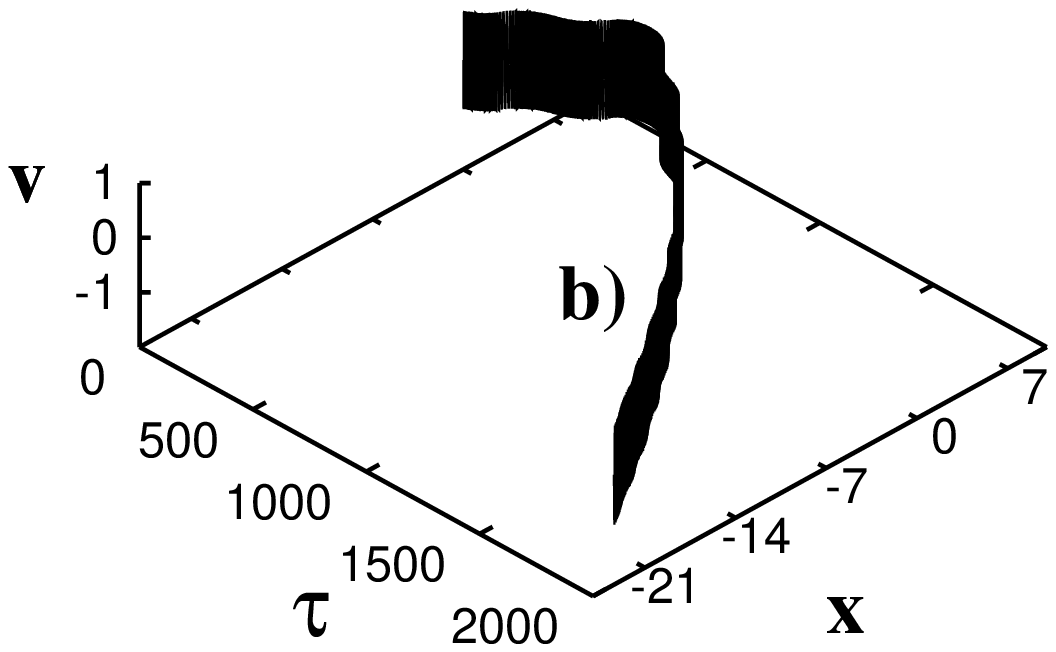}
\includegraphics[width=0.3\textwidth]{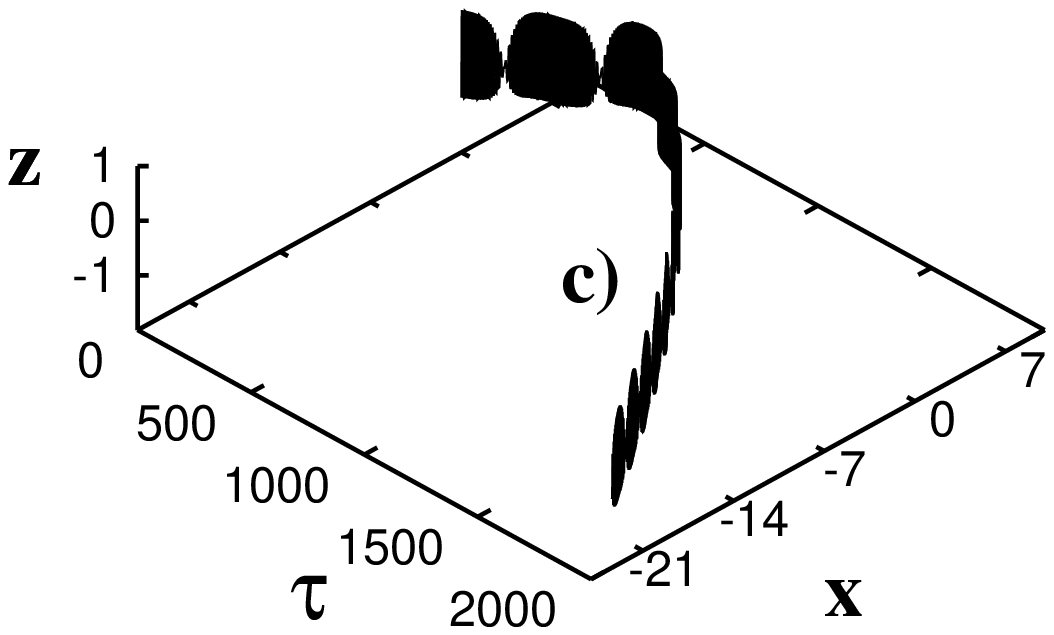}
\caption{The same as in Fig.~\ref{fig2} but far from the 
resonance ($\Delta=1$).} 
\label{fig3}
\end{figure}

\begin{figure}[!tpb]
\includegraphics[width=0.3\textwidth]{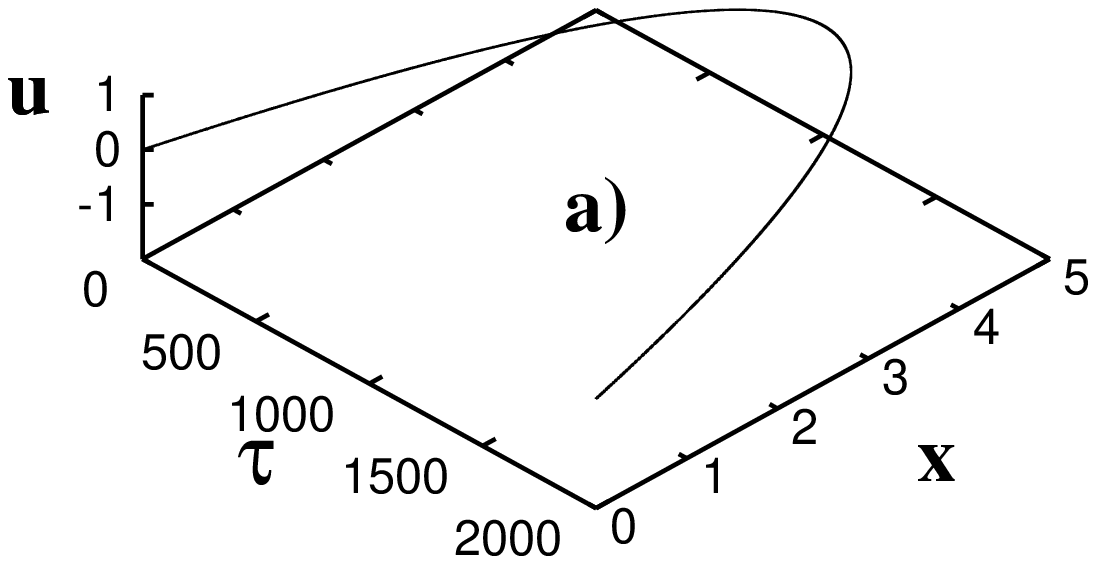}
\includegraphics[width=0.3\textwidth]{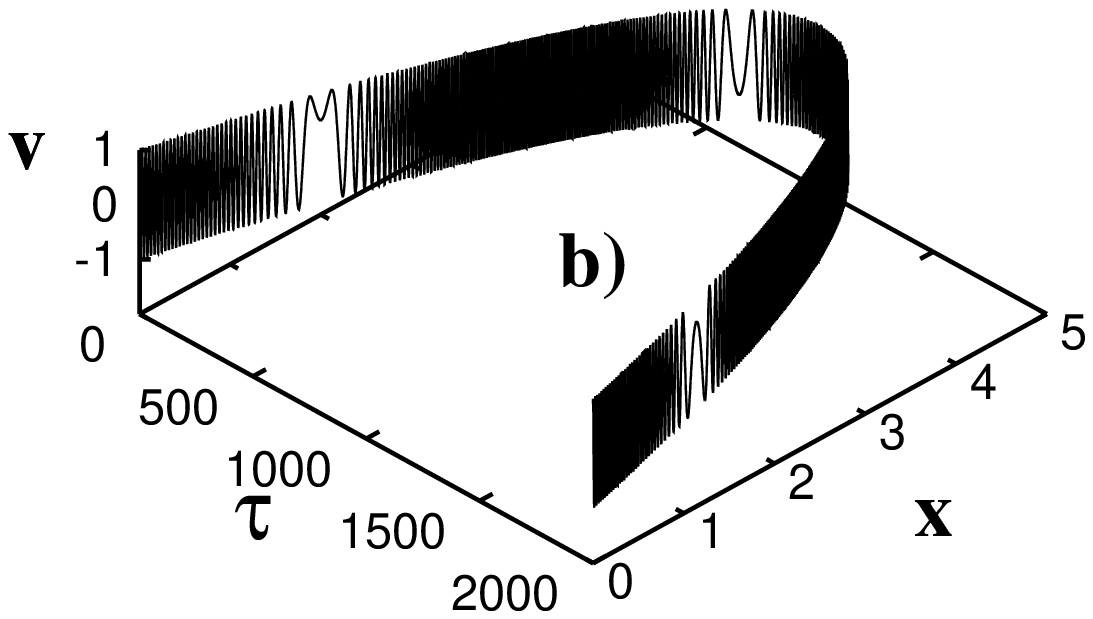}
\includegraphics[width=0.3\textwidth]{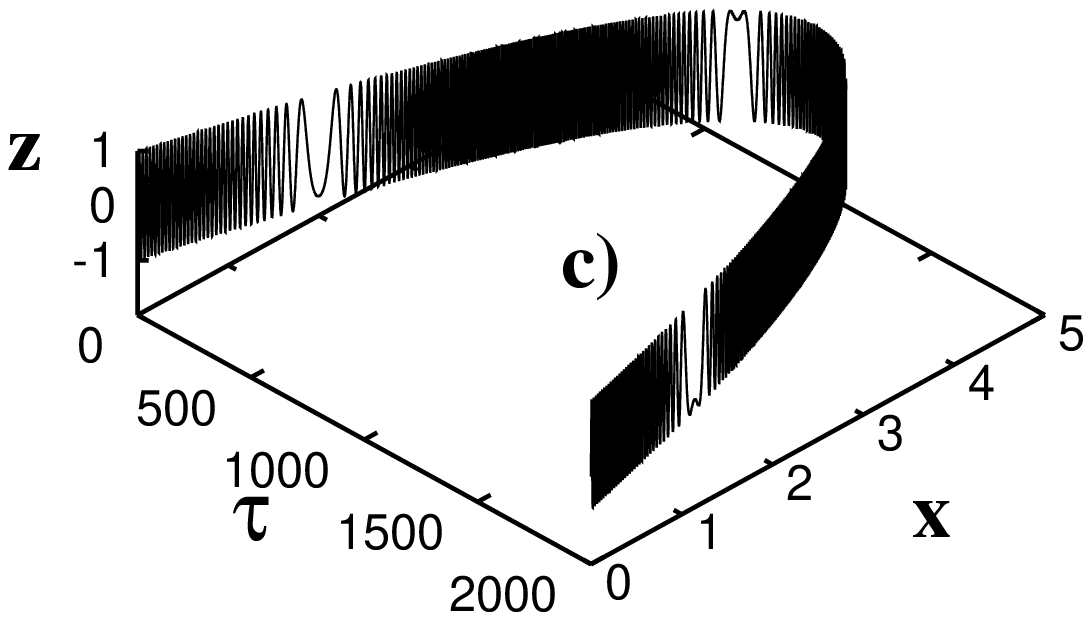}
\caption{The same as in Fig.~\ref{fig2} but at the resonance ($\Delta=0$).}  
\label{fig4}
\end{figure}

What is the ultimate reason of chaotic walking? For an optical lattice without an external force, 
it has been found in Ref.~\cite{PRA07} that instability is caused by 
the specific behavior of the Bloch-vector component  
of a moving atom, $u$, whose shallow oscillations between the standing-wave 
nodes are interrupted by 
sudden jumps with different amplitudes while atom crosses each node of the 
wave. It looks like a random like shots happened in a fully 
deterministic environment. The reason of chaotic walking in a tilted potential is 
the same. It follows from the second equation in the set ~(\ref{Hamsys}) that 
those jumps 
result in sudden changes of the atomic momentum while crossing 
nodes. If the value of the atomic energy is close to the separatrix 
one, the atom after the corresponding jump-like change in $p$ can either 
overcome the potential barrier and leave a potential well or  
it will be trapped by the well, or it will move as before crossing the node. 
The evolution of all the Bloch components in the regime of chaotic walking 
is shown in Fig.~\ref{fig2}. For comparison, we show in Figs.~\ref{fig3} and 
~\ref{fig4} their evolution in the regular regimes, far off the resonance and 
at exact resonance, respectively.  

\section{Fractal scattering of atoms in a tilted potential}
\pst

Different types of fractal-like structures may arise in chaotic
Hamiltonian systems (see reviews \cite{Gaspard,A09}). 
It is known from many studies in celestial mechanics 
\cite{PH86}, fluid dynamics \cite{PhysD,JETP04}, atomic physics 
\cite{Kolovsky,JETP03,JRLR06,PRA07}, cavity quantum electrodynamics  
\cite{PLA03,PU06}, underwater acoustics \cite{Chaos04} 
and other disciplines \cite{E88} that under certain conditions the motion inside an  
interaction region may have features that are typical for dynamical chaos, 
(homoclinic and heteroclinic tangles, fractals, strange invariant sets, 
positive finite-time Lyapunov exponents, etc.) although the particle's  
trajectories are not chaotic in a rigorous sense because chaos is defined as 
an irregular motion over infinite time. 

\begin{figure}[!tpb]
\includegraphics[width=0.3\textwidth]{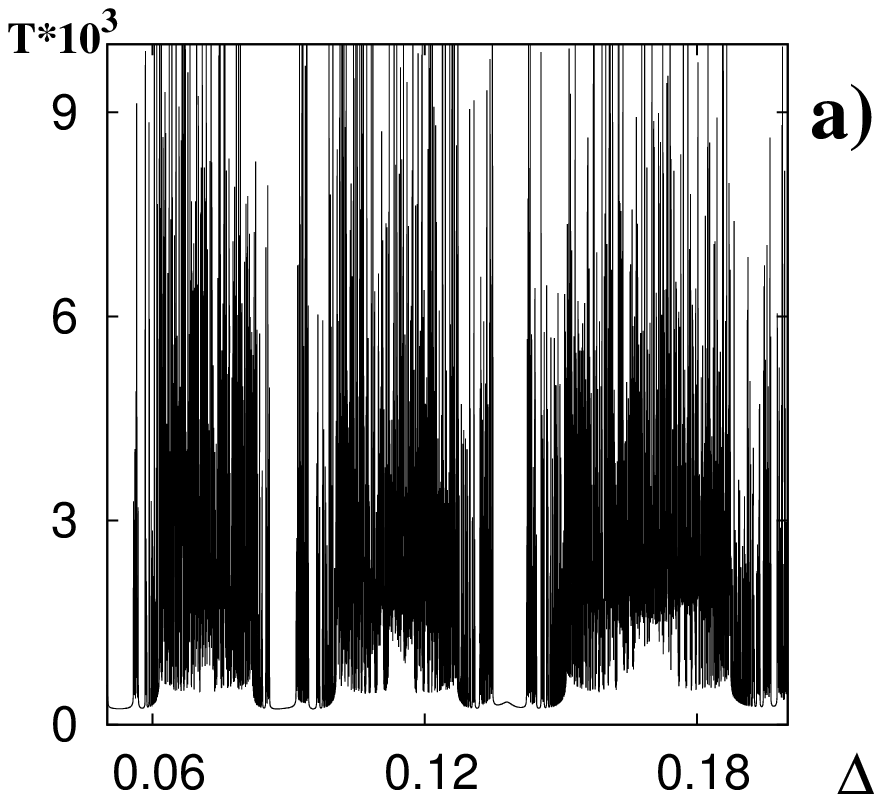}
\includegraphics[width=0.3\textwidth]{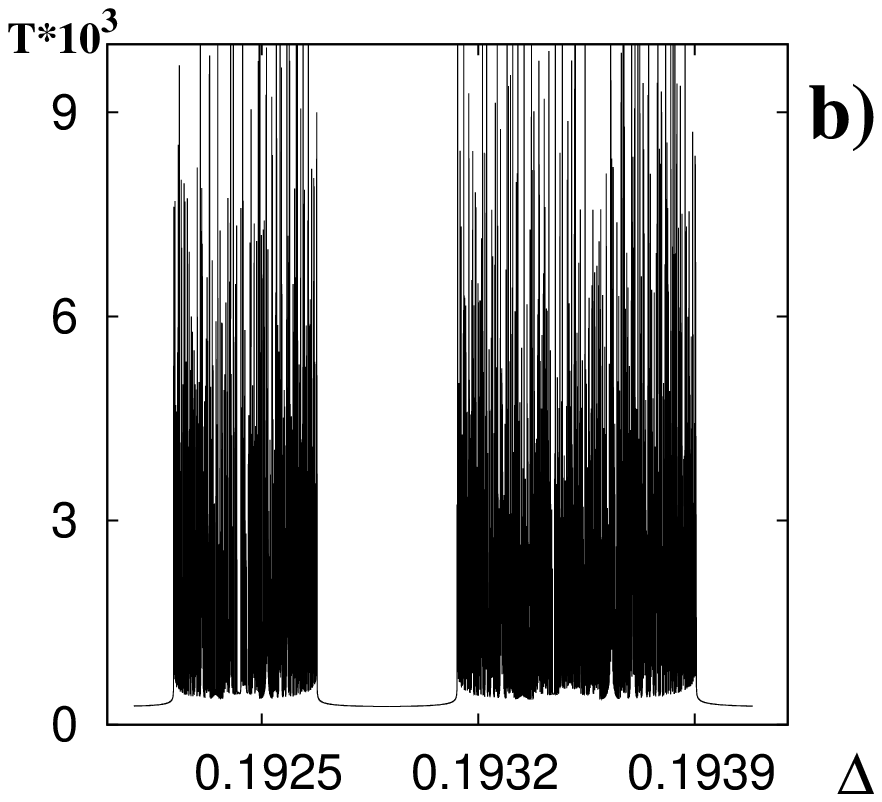}
\includegraphics[width=0.3\textwidth]{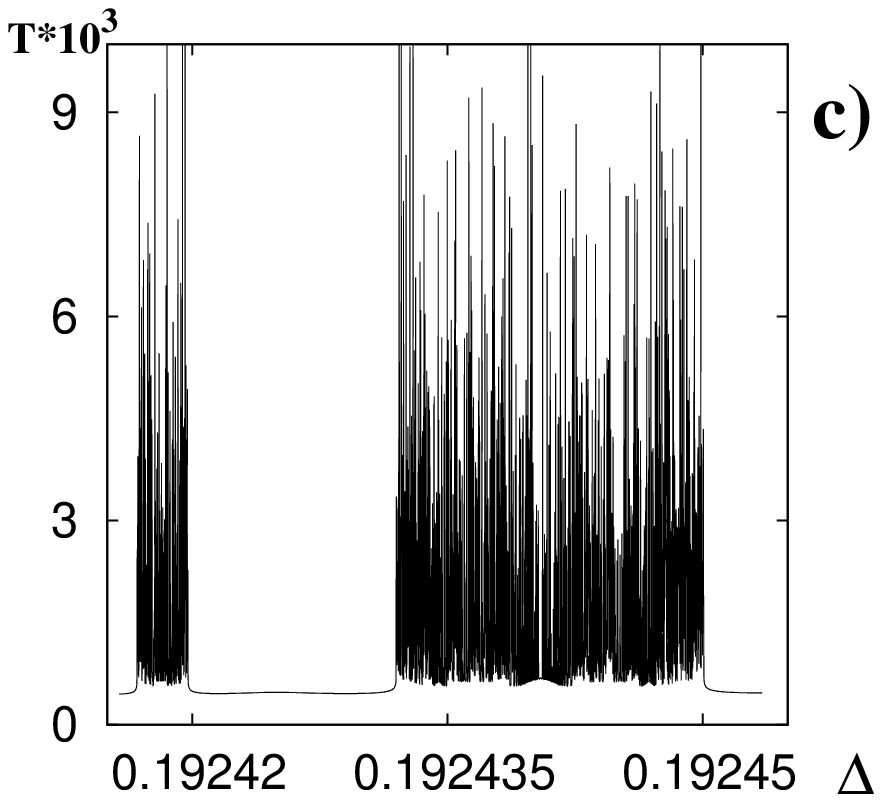}
\caption{Atomic dynamical fractal. Self-similar dependence of the exit time, 
$T$, with given initial position, $x_0=0$ and momentum $p_0=10$, 
on the detuning. The successive magnifications are shown.}  
\label{fig5}                                         
\end{figure}

Let us place atoms one by one at the point $x_0=0$
with the same value of the initial momentum $p_0 =10$ but 
change slightly the value of the detuning $\Delta$. All the other 
initial conditions and the control parameters are supposed to be the same 
for all the atoms.  
We fix the time moment $T$ when each atom crosses the point $x=0$ moving 
in the negative direction. The exit time function $T(\Delta)$ 
in Fig.~\ref{fig5} demonstrates the complicated structure with smooth 
intervals alternating with wildly oscillating peaks that cannot be resolved in principle, no
matter how large the magnification factor. The panels (b) and (c) in
Fig.~\ref{fig5} are successive 50 times magnifications of the detuning   
intervals shown in the panel (a). Further magnifications reveal a self-similar 
fractal-like structure that
is typical for Hamiltonian systems with chaotic scattering.
The exit time $T$ increases with increasing the magnification factor. 
The same picture is observed when computing the exit time function  
in the position and momentum spaces. It is a clear demonstration of a fractal-like 
behavior of chaotically walking atoms. 

It is established in theory of one and half degree-of-freedom systems 
that transient Hamiltonian chaos in the interaction region occurs due to 
existence of, at least, one non-attractive 
chaotic invariant set consisting of an infinite number of localized unstable 
periodic orbits and aperiodic orbits. This set possesses stable and unstable 
manifolds extending into the regions of regular motion. The particles with the 
initial positions close to the stable manifold follow the chaotic-set 
trajectories for a comparatively long time, then deviate 
from them, and leave the interaction region along the unstable manifold. 
In a typical Hamiltonian system there exists an infinite number of trajectories of zero measure 
with infinite exit time which belong to that chaotic invariant set.  
Our system with two and half degrees of freedom is a much more complicated one, and it is practically 
impossible to reveal the corresponding chaotic invariant set with its  
stable and unstable manifolds. However, the mechanism of chaotic scattering 
and fractal-like structures should be the same. 

\begin{figure}[!tpb]
\includegraphics[width=0.5\textwidth]{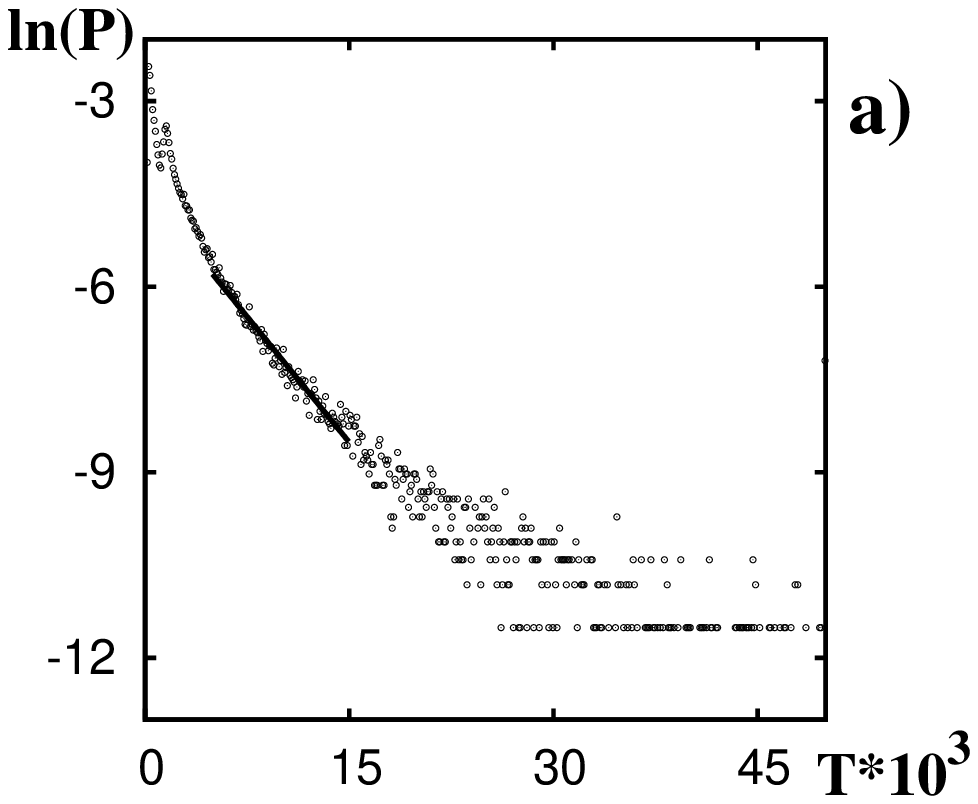}
\includegraphics[width=0.5\textwidth]{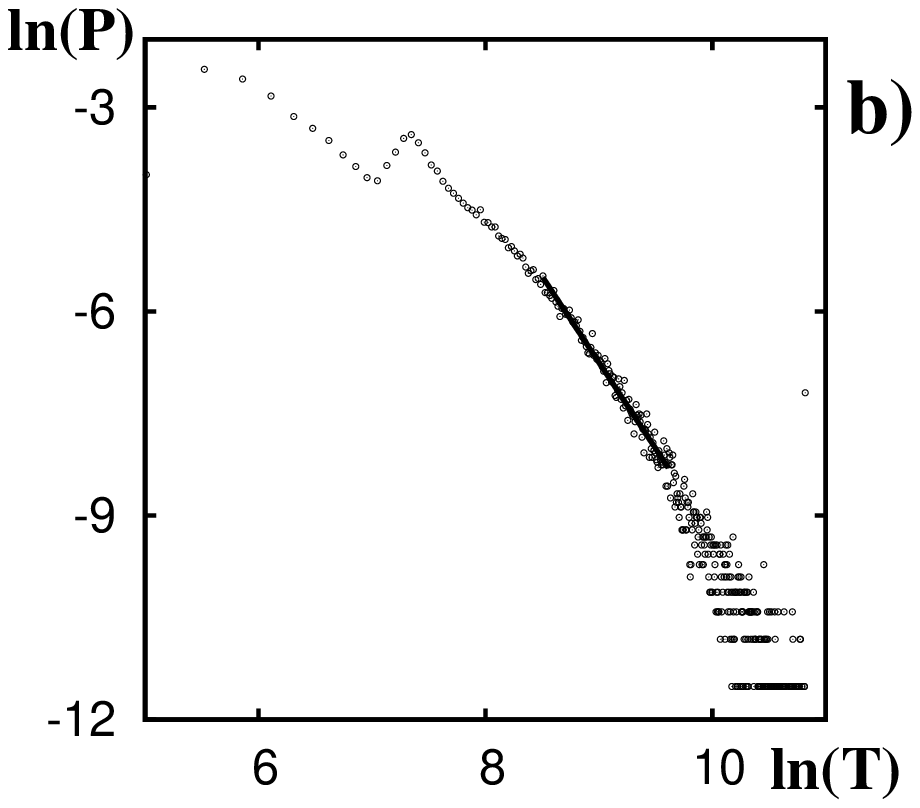}
\caption{The probability distribution function for exit times $T$  
in (a) semilogarithmic scale (exponential decay in the middle part with 
the exponent $\alpha=-0.000270722$) and (b) logarithmic scale 
(power-law decay at the tail with the coefficient  $\gamma=-2.53086$).}  
\label{fig6}
\end{figure}

The statistics of exit times $T$ is shown in Fig.~\ref{fig6} 
in a semilogarithmic and logarithmic scales. 
The probability distribution function (PDF) in this figure gives   
the probability for an atom to have a given value of $T$.
The bold straight line in Fig.~\ref{fig6}a implies that the PDF 
is exponential in its middle part, $P \sim \exp(-\alpha T)$, with the exponent  
$\alpha=-0.000270722$. However, the tail of the PDF is not exponential. 
To prove that we plot the function in the logarithmic scale 
in Fig.~\ref{fig6}b and compute the slope at the tail. It has no sense 
to calculate the slope at the very tail because of a small number of events with very large 
values of $T$. The bold straight line implies that 
the PDF is a power-law one, $P \sim  T^{-\gamma}$, 
with the coefficient  $\gamma=-2.53086$. It is interesting that the 
slope at the PDF tail around the value $-2.5$ is rather typical for many chaotic 
Hamiltonian systems \cite{Zas,Chaos06}. The reason of that is unclear. 

In hyperbolic chaotic systems the PDFs should decay  
exponentially because the phase space of such systems 
is homogeneous,  
and all the trajectories are unstable. It is not the case even with  
one and half degree-of-freedom systems with inhomogeneous phase space,  
where exist so-called stability islands embedded in a stochastic sea, 
because the borders of those islands are ``sticky''. It means that 
a typical chaotic 
trajectory, wandering in the stochastic sea, approaches the island's borders 
and ``stick'' to them for a long time. By that reason, the corresponding 
PDFs are not exponential but power-law ones 
at their tails. PDFs with power-law decay  
imply that the corresponding quantity, the exit time in our case, is scale 
invariant i.e., there is no a single dominant scale in the process. Geometrically 
it means that chaotic trajectories for such a process are self-similar.

\section{Conclusion}
\pst

It is shown that point-like atoms in a tilted optical potential
with a constant external force applied  can move chaotically
changing the direction of motion in a random-like way. The existence 
of chaos is confirmed by direct computation of the maximal finite-time 
Lyapunov exponent of the equations of motion 
that is shown to be positive in a range of the atom-laser detuning and the 
applied-force strength. The ultimate reason of  chaotic walking 
is the specific behavior of the Bloch-vector component  
of a moving atom, $u$, whose shallow oscillations between the standing-wave 
nodes are interrupted by sudden jumps with different amplitudes while atom crosses each node of the 
wave. It is demonstrated numerically that such a behavior arises exactly at 
those values of the detuning for which the Lyapunov exponent is positive 
and atoms move chaotically. We illustrate different regimes of the center-of-mass 
motion simply varying the detuning. It is an easy way to manipulate  
the atomic transport in tilted optical lattices. 

Treating motion of atoms in a tilted optical lattice as a scattering problem, we 
show that the scattering of atoms under conditions of chaotic walking 
is chaotic and typical for Hamiltonian systems. Fixing the time moment $T$ 
when atoms with slightly different values of the detuning, momentum or initial position
cross a fixed point ($x=0$), we show that the corresponding scattering functions  
demonstrate the complicated structure that cannot be resolved in principle, no
matter how large the magnification factor. Owing to that the probability 
to have a given value of $T$ is not exponential but decays at its tail 
by a power law.

\section*{Acknowledgments} 
This work was supported  by the Integration grant from the Far-Eastern 
and Siberian branches of the Russian Academy of Sciences (12-II-0-02-001),  
and by the Program ``Fundamental Problems of  Nonlinear Dynamics in Mathematics 
and Physics''. 

%
%
%

\end{document}